\documentclass[aps,prd,preprintnumbers,superscriptaddress,nofootinbib,notitlepage,twocolumn]{revtex4-2}
\usepackage[pdftex]{graphicx}
\usepackage{bm,latexsym,amsmath,amssymb,amsfonts,mathrsfs}
\usepackage{color}
\allowdisplaybreaks[1]
\usepackage[pdftex,colorlinks=true,linkcolor=blue,citecolor=cyan]{hyperref}
\newcommand*{\D}{{\rm d}}
\newcommand*{\mpl}{M_{\rm Pl}}
\begin{document}
%-----------------------------------------------------------------%
\title{Perturbations and quasi-normal modes of black holes with time-dependent scalar hair in shift-symmetric scalar-tensor theories}
%-----------------------------------------------------------------%
\author{Keitaro~Tomikawa}
\email[Email: ]{k.tomikawa@rikkyo.ac.jp}
\affiliation{Department of Physics, Rikkyo University, Toshima, Tokyo 171-8501, Japan
}
\author{Tsutomu~Kobayashi}
\email[Email: ]{tsutomu@rikkyo.ac.jp}
\affiliation{Department of Physics, Rikkyo University, Toshima, Tokyo 171-8501, Japan
}
%-----------------------------------------------------------%
\begin{abstract}
We study odd parity perturbations of spherically symmetric black holes
with time-dependent scalar hair in shift-symmetric higher-order scalar-tensor theories.
The analysis is performed in a general way without assuming
the degeneracy conditions. Nevertheless, we end up with second-order equations
for a single master variable, similarly to cosmological tensor modes. We thus identify
the general form of the quadratic Lagrangian for the odd parity perturbations,
leading to a generalization of the Regge-Wheeler equation.
We also investigate the structure of the effective metric for the master variable
and refine the stability conditions.
As an application of our generalized Regge-Wheeler equation,
we compute the quasi-normal modes of a certain nontrivial black hole solution.
Finally, our result is extended to include the matter energy-momentum tensor as
a source term. 
\end{abstract}
%-----------------------------------------------------------------%
\preprint{RUP-21-3}
\maketitle
%-----------------------------------------------------------------%
\section{Introduction}

The remarkable first direct detection of gravitational waves from
a binary black hole merger~\cite{Abbott:2016blz}
has opened a new era of astrophysics and gravitational physics.
Gravitational waves are becoming more and more important
as a probe of strong gravitational fields and
as a tool for testing gravity in the strong field regime.
To establish general relativity on a firmer basis,
it is necessary to study the predictions from alternative theories
such as scalar-tensor theories
and test them against gravitational wave observations.
In modified theories gravity,
black holes may have hair, i.e., nontrivial configurations of
scalar or other extra degrees of freedom around themselves,
and the perturbation dynamics may also differ from
that in general relativity.
Therefore, identifying the general action for
perturbations around a hairy black hole will help
to achieve the above purpose.
The results will be useful for instance for the computation of
quasi-normal modes (QNMs) in modified gravity.

In this paper,
we determine the general action governing odd parity perturbations
around a spherically symmetric black hole dressed with a linearly time-dependent
scalar field.
To do so, we start from a covariant action for shift-symmetric
higher-order scalar-tensor theories admitting such time-dependent
scalar hair and second-order field equations at least in the odd parity sector.
In contrast to Refs.~\cite{Kase:2014baa,Franciolini:2018uyq,Kuntz:2020yow},
we do not take the effective-field-theory (EFT) approach,
because our background scalar-field configuration depends not only on
the radial coordinate but also on time,
which breaks the usual assumption of the EFT on the symmetry.
In such a case, it is probably more convenient to start from a covariant action.

To make the results as general as possible, we work with general
shift-symmetric higher-order scalar-tensor theories whose action 
depends on the curvature tensors and
first and second derivatives of the scalar field
in such a way that yields healthy second-order field equations
for gravitational-wave degrees of freedom.
Probably the most well-known example of such theories is
the Horndeski theory~\cite{Horndeski:1974wa}, or, equivalently,
the generalized Galileon theory~\cite{Deffayet:2011gz,Kobayashi:2011nu}.
The perturbation theory for spherically symmetric black holes
with static hair has been developed in Refs.~\cite{Kobayashi:2012kh,Kobayashi:2014wsa}.
When restricted to the shift-symmetric subclass, the Horndeski theory
admits linearly time-dependent scalar hair, as was first
demonstrated in Ref.~\cite{Babichev:2013cya}
and later generalized in Ref.~\cite{Kobayashi:2014eva},
thus evading the no-hair theorem in shift-symmetric scalar-tensor theories~\cite{Hui:2012qt}.
The black-hole perturbation theory can be extended to
the case with linearly time-dependent scalar hair~\cite{Ogawa:2015pea,Takahashi:2016dnv}
(see Ref.~\cite{Khoury:2020aya} for a recent update).
Though the Horndeski theory is the most general scalar-tensor theory
having second-order field equations both for the metric and scalar field,
later it was noticed that it can further be generalized while
maintaining one scalar and two tensorial degrees
of freedom~\cite{Gleyzes:2014dya,Langlois:2015cwa,Crisostomi:2016czh,BenAchour:2016fzp}
(see also Ref.~\cite{Zumalacarregui:2013pma} for an earlier work
seeking theories beyond Horndeski by means of a disformal transformation
of the metric).
The basic idea behind this generalization is that if some of the
field equations are degenerate, then the number of dynamical degrees of freedom
is reduced and thus the system can retain one scalar and two tensorial degrees of
freedom even if the field equations are apparently of higher order.
Theories with such a structure are called
degenerate higher-order scalar-tensor (DHOST) theories.
See Refs.~\cite{Langlois:2017mdk,Langlois:2018dxi,Kobayashi:2019hrl}
for a review. In DHOST theories in which second derivatives of the scalar field
appear quadratically in the Lagrangian (i.e., quadratic DHOST theories),
odd parity perturbations of spherically symmetric black holes
with linearly time-dependent scalar hair have been studied in Ref.~\cite{Takahashi:2019oxz},
with the results showing that the master variable for the odd parity perturbations
obeys a second-order equation having essentially the same structure
as that in the case of the Horndeski theory~\cite{Ogawa:2015pea,Takahashi:2016dnv}.
More recently, it was argued that the dangerous ghost degrees of freedom
remain to be absent even if one relaxes the degeneracy conditions
so that the system is degenerate only in the unitary gauge,
giving rise to the notion of ``U-degenerate'' theories~\cite{DeFelice:2018ewo}.
If a theory is treated as a low-energy effective theory rather than
a complete one, it is sufficient to require that
no ghost degrees of freedom emerge within the
regime of validity of the effective theory. This viewpoint allows us to consider
the theories in which the degeneracy conditions are detuned
slightly~\cite{Motohashi:2019ymr}.
Detuning the degeneracy conditions help to resolve the problem of
infinite strong coupling in the even parity sector~\cite{Minamitsuji:2018vuw,deRham:2019gha}.

In light of these developments,
we will study black-hole perturbations in higher-order scalar-tensor theories
that are most closely related to cubic DHOST theories~\cite{BenAchour:2016fzp},
but without imposing the degeneracy conditions.
Still, at least in the odd parity sector,
we will have a second order equation for a single master variable
and can thus determine the general form of the action for the master variable,
generalizing the previous results~\cite{Ogawa:2015pea,Takahashi:2016dnv,Takahashi:2019oxz}.

This paper is organized as follows.
In the next section,
we present the covariant action for scalar-tensor theories which we will work with.
In Sec.~III we give an example of spherically symmetric background solutions
with time-dependent scalar hair. Then, in Sec.~IV we determine the
general action for odd parity perturbations and derive the generalized
Regge-Wheeler equation. We also refine the previous notion of the stability conditions
by investigating the structure of the effective metric for gravitons.
In Sec.~V, we calculate the QNMs
of the black hole solution we present in Sec.~III.
Section~VI is devoted to a summary of conclusions.
In the appendices we argue the generality of our action for
odd parity perturbations. We also generalize the Regge-Wheeler equation
derived in the main text to include the matter energy-momentum tensor
as a source term.

\section{Higher-Order Scalar-Tensor Theories}

We consider a system composed of the metric $g_{\mu\nu}$ and 
the scalar field $\phi$
described by the action~\cite{Langlois:2015cwa,Crisostomi:2016czh,BenAchour:2016fzp}
\begin{align}
    &S_{\rm grav}=\int\D^4x\sqrt{-g}\biggl[
    F_0(X)+F_1(X)\Box\phi+F_2(X) R
    \notag \\ &
    +\sum_{I=1}^5A_I(X)L^{(2)}_{I}
    +F_3(X)G_{\mu\nu}\phi^{\mu\nu}
    +\sum_{I=1}^{10}B_I(X)L_I^{(3)}
    \biggr], \label{eq:action1}
\end{align}
where $X:=-\phi_\mu\phi^\mu/2$,
$\phi_\mu:=\nabla_\mu\phi$, $\phi_{\mu\nu}=\nabla_\nu\nabla_\mu\phi$,
$R$ is the Ricci scalar, and $G_{\mu\nu}$ is the Einstein tensor.
Here, $L_I^{(2)}$ are quadratic in
the second derivatives of the scalar field and are written explicitly as
\begin{align}
    &L_1^{(2)}=\phi_{\mu\nu}\phi^{\mu\nu},\quad
    L_2^{(2)}=(\Box\phi)^2,\quad
    L_3^{(2)}=(\Box \phi)\phi^{\mu} \phi_{\mu \nu}\phi^{\nu},
    \notag \\ &
    L_4^{(2)}=\phi^{\mu}\phi_{\mu \rho}\phi^{\rho \nu}\phi_{\nu},\quad
    L_{5}^{(2)}=(\phi^{\mu}\phi_{\mu \nu}\phi^{\nu})^{2}.
\end{align}
Similarly, $L_I^{(3)}$ are cubic in the second derivatives of
the scalar field and are given by
\begin{align}
    &L_{1}^{(3)} = (\Box \phi)^{3},&
    &L_{2}^{(3)} = (\Box \phi)\phi_{\mu \nu}\phi^{\mu \nu},& \nonumber\\
    &L_{3}^{(3)} = \phi_{\mu \nu} \phi^{\nu \rho} \phi^{\mu}_{\rho},& 
    &L_{4}^{(3)} = (\Box \phi)^{2} \phi_{\mu} \phi^{\mu \nu} \phi_{\nu},& \nonumber\\
    &L_{5}^{(3)} = \Box \phi \phi_{\mu} \phi^{\mu \nu} \phi_{\nu \rho}\phi^{\rho},&
    &L_{6}^{(3)} = \phi_{\mu \nu} \phi^{\mu \nu} \phi_{\rho} \phi^{\rho \sigma} \phi_{\sigma},& \nonumber \\
    &L_{7}^{(3)} = \phi_{\mu} \phi^{\mu \nu} \phi_{\nu \rho} \phi^{\rho \sigma} \phi_{\sigma},&
    &L_{8}^{(3)} = \phi_{\mu} \phi^{\mu \nu} \phi_{\nu \rho} \phi^{\rho} \phi_{\sigma} \phi^{\sigma \lambda} \phi_{\lambda}, & \quad \nonumber \\
    &L_{9}^{(3)} = \Box \phi (\phi_{\mu} \phi^{\mu \nu} \phi_{\nu})^{2},&
    &L_{10}^{(3)} = (\phi_{\mu} \phi^{\mu \nu} \phi_{\nu})^{3}. 
\end{align}
These exhaust possible terms built from $\phi_\mu$ and $\phi_{\mu\nu}$
and quadratic/cubic in $\phi_{\mu\nu}$.
The functions $F_0$, $F_1$, $F_2$, $F_3$, $A_I$, and $B_I$
depend only on $X$, so that the theory has shift symmetry,
$\phi\to\phi+c$.

In general, the action~\eqref{eq:action1} yields
higher-order equations of motion for the metric and the scalar field,
resulting in the dangerous Ostrogradsky ghost.
One can circumvent this by imposing
the degeneracy conditions among the functions
$F_2$, $F_3$, $A_I$, and $B_I$~\cite{Langlois:2015cwa,Crisostomi:2016czh,BenAchour:2016fzp}.
In such degenerate theories, one arrives in the end at
a set of second-order equations by combining the different components of field equations,
and thus can remove the unstable ghost degrees of freedom.
One may relax the degeneracy conditions so that
the theory is degenerate at least in the unitary gauge,
which can still provide a healthy class of theories
called U-degenerate theories~\cite{DeFelice:2018ewo}.
If the action~\eqref{eq:action1} is regarded
as a low-energy truncation of some complete theory,
detuning the degeneracy conditions is acceptable because
a ghost degree of freedom itself is not problematic
from the effective-field-theory viewpoint~\cite{Motohashi:2019ymr}.

In this paper, we do not assume any particular relations
among the functions in the action.
Nevertheless, we can handle the relevant equations
and derive the universal form of the
quadratic Lagrangian for odd mode perturbations
around a spherically symmetric background
with time-dependent scalar hair.

\section{Spherically Symmetric Background}\label{sec:Background}

Let us start with a background solution.
We consider static and spherically symmetric spacetime
whose metric is of the form
\begin{align}
    \D s^2=-A(r)\D t^2+\frac{\D r^2}{B(r)}+r^2C(r)
    \D\sigma^2,
    %\left(\D\theta^2+\sin^2\theta\D\varphi^2\right).
    \label{eq:bg_metric}
\end{align}
where $\D\sigma^2:=\D\theta^2+\sin^2\theta\D\varphi^2$.
At this point we introduce $C(r)$ to reproduce all the
relevant field equations from the action principle.
After deriving the field equations one may put $C(r)=1$
by redefining the radial coordinate (see, e.g.,~\cite{Motohashi:2016prk}).

The scalar field is assumed to be dependent linearly on
the time coordinate,
\begin{align}
    \phi(t,r)=\mu t +\psi(r), \label{eq:bg_scalar}
\end{align}
where $\mu$ is a constant.
Without loss of generality we assume that $\mu>0$.
This configuration is consistent with the static metric~\eqref{eq:bg_metric}
because the action~\eqref{eq:action1} depends on $\phi$
only through its derivatives.

The crucial points of the ansatz~\eqref{eq:bg_scalar}
are the following. First, by assuming a linearly time-dependent
scalar field one can avoid the postulate
in the no-hair theorem of~\cite{Hui:2012qt},
which makes it easier to obtain hairy solutions.
Indeed, spherically symmetric
black hole solutions with such a scalar field configuration
have been found
in the context of the Horndeski theory~\cite{Babichev:2013cya,Kobayashi:2014eva,Babichev:2016rlq,Babichev:2016fbg,Emond:2019myx}
and beyond-Horndeski/
DHOST theories~\cite{Babichev:2016kdt,Babichev:2017guv,Motohashi:2019sen,Minamitsuji:2019shy,Takahashi:2020hso}.
Second, it has been assumed in the formulation of
the EFT of black hole
perturbations~\cite{Franciolini:2018uyq,Kuntz:2020yow}
that the scalar field depends only on the radial coordinate.
Therefore, for a time-dependent scalar field configuration
the previous result from the effective field theory approach
cannot be used straightforwardly, and it is interesting
to explore a general form of the effective action for
black hole perturbations in the presence of
time-dependent hair.

Substituting the metric~\eqref{eq:bg_metric} and
the scalar field ansatz~\eqref{eq:bg_scalar} to
the action~\eqref{eq:action1} and varying it
with respect to $A$, $B$, $C$, and $\psi$,
one is able to derive the background field equations.
We write the resultant field equations as
${\cal E}_A=0$, ${\cal E}_B=0$, ${\cal E}_C=0$, and ${\cal E}_\psi=0$,
whose explicit expressions
are not important in the present paper.
These equations do not reduce to
second-order differential equations in general,
because we do not impose any degeneracy conditions.
However, as far as the odd mode perturbations are concerned,
we do not need to care about the higher-order nature of the
background equations.
We will just use (some of) these background equations
in their original form
to simplify the quadratic action for
the odd mode perturbations,
whether they are of second order or higher.

Before proceeding to the analysis of perturbations,
let us present a simple explicit example of background solutions.
An interesting class of solutions often studied in the literature
is a stealth Schwarzschild black hole with $X=X_0=\;$const.
One can see that our field equations admit the solution
\begin{align}
    A=B=1-\frac{r_h}{r},\quad X=X_0=\frac{\mu^2}{2},
\end{align}
provided that the functions in the action~\eqref{eq:action1}
satisfy the following equations (cf.~\cite{Minamitsuji:2019shy}):
\begin{align}
    &F_0(X_0)=0,\quad F_{0X}(X_0)=0,\quad F_{1X}(X_0)=0
    \notag \\ &
    A_1(X_0)+A_2(X_0)=0,\quad A_{1X}(X_0)+A_{2X}(X_0)=0,
    \notag \\ &
    B_2(X_0)=-\frac{1}{2}B_3(X_0)=9B_1(X_0),
    \notag \\ &
    B_4(X_0)+B_6(X_0)-B_{1X}(X_0)-B_{2X}(X_0)-\frac{5}{9}B_{3X}(X_0)
    \notag \\ &
    =\frac{6}{X_0}B_1(X_0).\label{bg_soln_rels}
\end{align}
Note that these relations are compatible with
the degeneracy conditions in the class $^2$N-I $+$ $^3$M-I
degenerate theories in the terminology of~\cite{BenAchour:2016fzp},
and therefore the above solution is admitted even if one
concentrates on a degenerate theory.
From $2X=\mu^2=\mu^2/A-B(\D\psi/\D r)^2$ we have 
\begin{align}
    \psi=\pm\mu\left[
    2\sqrt{r_hr}+r_h\ln\left(\frac{\sqrt{r}-\sqrt{r_h}}{\sqrt{r}+\sqrt{r_h}}\right)
    \right].\label{eq:psi-null}
\end{align}
We choose the ``$+$'' branch because we have
$\phi\simeq \mu [t\pm r_h\ln(r/r_h-1) ]+\;$const
near the horizon and it is regular at the horizon only in the ``$+$'' branch,
as is clear by expressing $\phi$ in terms of the ingoing null coordinate 
$v=t+r+r_h\ln(r/r_h-1)$~\cite{Babichev:2013cya}.

\section{Odd Parity Perturbations}

\subsection{Derivation of the Quadratic Lagrangian and the Effective Metric}

Let us consider the odd mode metric perturbations,
\begin{align}
    g_{\mu\nu}=\overline{g}_{\mu\nu}+h_{\mu\nu},
\end{align}
where $\overline{g}_{\mu\nu}$ is the background metric~\eqref{eq:bg_metric}
with $C(r)=1$.
The scalar field does not have an odd mode perturbation.
Among the ten components, $h_{ta}$, $h_{ra}$, and $h_{ab}$ are concerned with
odd parity modes, where $a=\theta, \varphi$.
Using the spherical harmonics $Y_{\ell m}(\theta,\varphi)$,
we follow the standard procedure and
expand the odd mode perturbations as
\begin{align}
    h_{t\theta}&=-\frac{1}{\sin\theta}\partial_\varphi
    \sum_{\ell=2}^\infty\sum_{m=-\ell}^{\ell} h_0^{(\ell m)}
    (t,r)Y_{\ell m}(\theta,\varphi),\label{eq:pert_def_1}
    \\
    h_{t\varphi}&=\sin\theta\partial_\theta
    \sum_{\ell=2}^\infty\sum_{m=-\ell}^{\ell} h_0^{(\ell m)}
    (t,r)Y_{\ell m}(\theta,\varphi),
    \\
    h_{r\theta}&=-\frac{1}{\sin\theta}\partial_\varphi
    \sum_{\ell=2}^\infty\sum_{m=-\ell}^{\ell} h_1^{(\ell m)}
    (t,r)Y_{\ell m}(\theta,\varphi),
    \\
    h_{r\varphi}&=\sin\theta\partial_\theta
    \sum_{\ell=2}^\infty\sum_{m=-\ell}^{\ell} h_1^{(\ell m)}
    (t,r)Y_{\ell m}(\theta,\varphi).\label{eq:pert_def_4}
\end{align}
The odd parity part of $h_{ab}$ can also be expressed using
a single pseudo-scalar function, say $h_2$, but we adopt the Regge-Wheeler gauge
in which $h_2=0$ and accordingly $h_{ab}=0$.

We substitute Eqs.~\eqref{eq:pert_def_1}--\eqref{eq:pert_def_4}
to the action~\eqref{eq:action1} and expand it to second order
in perturbations. In doing so
one can remove many terms by using the background equations.
Performing the angular integrations,
we arrive in the end at the general action
\begin{align}
S_{\rm grav}=\sum_{\ell=2}^\infty\sum_{m=-\ell}^\ell \int\D t \D r
{\cal L}_{\ell m}^{(2)},\label{gravaction2}
\end{align}
where, omitting the labels $(\ell m)$ from $h_0$ and $h_1$,
\begin{widetext}
\begin{align}
{\cal L}_{\ell m}^{(2)}&=\frac{1}{2}\left\{\left[
\frac{2}{r^2}(ra_3)'+a_1 \right]|h_0|^2+a_2|h_1|^2
+a_3\left(
|\dot h_1|^2-2\dot h_1^*h_0'+|h_0'|^2+\frac{4}{r}\dot h_1^*h_0
\right)+a_4h_1^*h_0\right\} +{\rm c.c.}.\label{eq:quadratic_Lagrangian}
\end{align}
The coefficients are given by
\begin{align}
a_1&=\frac{c_\ell}{2r^2\sqrt{AB}}
\left\{F_2+\frac{\mu^2}{A}A_1+\frac{B\psi'X'}{2}F_{3X}
+\frac{\mu^2 }{A\psi'}\left[
\frac{2B(\psi')^2}{r}-\frac{(AX)'}{A}\right]B_2
+\frac{3\mu^2B\psi'}{rA}B_3
-\frac{\mu^2B\psi'X'}{A}B_6
\right\},\label{coeff:a1}
\\
a_2&=-\frac{c_\ell}{2}\frac{\sqrt{AB}}{r^2}
\biggl\{F_2-B(\psi')^2A_1
-\frac{B\psi'X'}{2}F_{3X}
-B\psi'\left[\frac{2B(\psi')^2}{r}-\frac{(AX)'}{A}\right]B_2
\notag \\ & \quad\quad\quad\quad\quad\quad\quad
-\frac{3B^2(\psi')^3}{r}B_3+B^2(\psi')^3X'B_6
\biggr\},
\\
a_3&=\frac{\ell(\ell+1)}{2}\sqrt{\frac{B}{A}}
\biggl\{
F_2+2XA_1-\frac{B\psi'X'}{2}F_{3X}
+\frac{2X}{\psi'}\left[\frac{2B(\psi')^2}{r}-\frac{(AX)'}{A}\right]B_2
\notag \\ & \quad\quad\quad\quad\quad\quad\quad
+\frac{3X}{\psi'}\left[
\frac{B(\psi')^2}{r}-X\frac{A'}{A}-\frac{\mu^2X'}{2AX}
\right]B_3-2B\psi'XX'B_6
\biggr\},
\\
a_4&=-\frac{c_\ell}{r^2}\sqrt{\frac{B}{A}}\mu
\left\{\psi'A_1+\frac{X'}{2}F_{3X}
+\left[\frac{2B(\psi')^2}{r}-\frac{(AX)'}{A}\right]B_2
+\frac{3B(\psi')^2}{r}B_3-B(\psi')^2X'B_6
\right\},\label{coeff:a4}
\end{align}
\end{widetext}%
with $c_\ell=(\ell-1)\ell(\ell+1)(\ell+2)$.
Here a dot and a prime denote differentiation with respect to
$t$ and $r$, respectively. Following Ref.~\cite{Takahashi:2019oxz},
it is convenient to write these coefficients as
\begin{align}
&a_1=\frac{c_\ell}{4r^2\sqrt{AB}}{\cal F}(r),
\quad 
a_2=-\frac{c_\ell}{4}\frac{\sqrt{AB}}{r^2}{\cal G}(r),
\notag \\
&a_3=\frac{\ell(\ell+1)}{4}\sqrt{\frac{B}{A}}{\cal H}(r),
\quad 
a_4=\frac{c_\ell}{2r^2}\sqrt{\frac{B}{A}}{\cal J}(r) ,
\end{align}
where ${\cal F}$, ${\cal G}$, ${\cal H}$, and ${\cal J}$ have
a dimension of (mass)$^2$.
For the Schwarzschild solution in general relativity,
we simply have ${\cal F}={\cal G}={\cal H}=\mpl^2=(8\pi G)^{-1}$
and ${\cal J}=0$.

Before proceeding to further reduction of the Lagrangian~\eqref{eq:quadratic_Lagrangian},
let us point out two things, both of which come essentially from the
fact that the odd parity modes constitute a part of tensorial metric perturbations.
First, note that only
$F_2$, $F_3$, $A_1$, $B_2$, $B_3$, and $B_6$
participate in the above result.
Contributions from the other terms are dropped from the action
upon using the background equations.
This is as expected because it is known that only these terms contribute to
the tensor modes on a cosmological background~\cite{deRham:2016wji,Langlois:2017mxy}.
Second,
it should be emphasized that the quadratic Lagrangian~\eqref{eq:quadratic_Lagrangian}
is derived without using any degeneracy conditions.
This is also not surprising because tensorial metric perturbations
in the theory~\eqref{eq:action1}
obey second-order equations without regard to the degeneracy conditions.
Therefore, our result can be used, for example, to
U-degenerate theories~\cite{DeFelice:2018ewo}
and detuned (``scordatura'') DHOST theories~\cite{Motohashi:2019ymr}.

Now let us rewrite the Lagrangian~\eqref{eq:quadratic_Lagrangian}
in terms of a single master variable.
This can be done straightforwardly, following closely
Refs.~\cite{Ogawa:2015pea,Takahashi:2016dnv,Takahashi:2019oxz}.
First, we introduce an auxiliary field $\chi=\chi^{(\ell m)}(t,r)$ and
rewrite the Lagrangian~\eqref{eq:quadratic_Lagrangian}
in an equivalent way as
\begin{align}
    {\cal L}_{\ell m}^{(2)}
    &=\frac{1}{2}\biggl[a_1|h_0|^2+a_2|h_1|^2+a_4h_1^*h_0
    \notag \\ &\quad 
    +2a_3\chi^*\left(-\frac{1}{2}\chi+\dot h_1-h_0'+\frac{2}{r}h_0\right)
    \biggr]+{\rm c.c.}.\label{eq:Lagrangian-chi}
\end{align}
Variation with respect to $h_0^*$ and $h_1^*$ leads, respectively, to
\begin{align}
    a_1h_0+(a_3\chi)'+\frac{2a_3}{r}\chi +\frac{1}{2}a_4h_1&=0,\label{eq:h0eq}
    \\
    a_2h_1-a_3\dot \chi +\frac{1}{2}a_4h_0&=0,\label{eq:h1eq}
\end{align}
which can be solved for $h_0$ and $h_1$ to express them
in terms of $\chi$, $\dot\chi$, and $\chi'$:
\begin{align}
    h_0&=-\frac{8a_2a_3(\chi/r)+4a_2(a_3\chi)'+2a_3a_4\dot\chi}%
    {4a_1a_2-a_4^2},
    \label{eq:soln-h0}
    \\
    h_1&=\frac{4a_3a_4(\chi/r)+2a_4(a_3\chi)'+2a_1a_3\dot\chi}%
    {4a_1a_2-a_4^2}
    .\label{eq:soln-h1}
\end{align}
(Here we assumed that ${\cal FG}+(B/A){\cal J}^2\neq 0$.)
Substituting Eqs.~\eqref{eq:soln-h0} and~\eqref{eq:soln-h1}
back to Eq.~\eqref{eq:Lagrangian-chi}, we obtain
\begin{align}
    {\cal L}_{\ell m}^{(2)}&=\frac{\ell(\ell+1)r^2}{4(\ell-1)(\ell+2)}\sqrt{\frac{B}{A}}
    \biggl\{b_1|\dot\chi|^2-b_2|\chi'|^2+b_3\dot\chi^*\chi'
    \notag \\ & \quad 
    -\left[\frac{\ell(\ell+1)}{2}\frac{{\cal H}}{r^2}+\frac{V}{2}\right]|\chi|^2
    \biggr\}+{\rm c.c.},\label{eq:Lagrangian-chi2}
\end{align}
where
\begin{align}
    &b_1=\frac{{\cal F}}{2A}\cdot \frac{A{\cal H}^2}{A{\cal FG}+B{\cal J}^2},
    %\notag \\ 
    \quad 
    b_2=\frac{{\cal G}B}{2}\cdot \frac{A{\cal H}^2}{A{\cal FG}+B{\cal J}^2},
    \notag \\
    &b_3=\frac{B{\cal J}}{A}\cdot \frac{A{\cal H}^2}{A{\cal FG}+B{\cal J}^2},
\end{align}
and 
\begin{align}
    V=2{\cal H}\left[r^2
    b_2 \sqrt{\frac{B}{A}}\left(\frac{\sqrt{A/B}}{r^2{\cal H}}
    \right)'\right]'-\frac{2{\cal H}}{r^2}.
\end{align}
The equation of motion that follows from this Lagrangian is given by
\begin{align}
    &b_1\ddot\chi -\frac{\sqrt{A/B}}{r^2}
    \left(r^2\sqrt{\frac{B}{A}}b_2\chi'\right)'
    +\frac{b_3}{2}\dot\chi'
    \notag \\ &
    +\frac{\sqrt{A/B}}{2r^2}
    \left(r^2\sqrt{\frac{B}{A}}b_3\dot \chi\right)'
    +\left[\frac{\ell(\ell+1)}{2}\frac{{\cal H}}{r^2}+\frac{V}{2}\right]\chi
    =0.\label{eq:eom_chi_1}
\end{align}

At this stage, it can be seen from the Lagrangian~\eqref{eq:Lagrangian-chi2}
that we need to impose
\begin{align}
    {\cal H}>0,
\end{align}
as otherwise modes with large $\ell$ would have large negative energy
and make the system unstable quickly.

One notices that Eq.~\eqref{eq:eom_chi_1} can be written
in the form
\begin{align}
    {\cal H}\Omega^2 Z^{\mu\nu}D_\mu D_\nu\chi -V\chi =0,\label{eq:RWZ01}
\end{align}
where $Z^{\mu\nu}$ is the inverse of the effective metric $Z_{\mu\nu}$~\cite{Babichev:2018uiw},
\begin{align}
    Z_{\mu\nu}\D x^\mu\D x^\nu&=
    \Omega^2\biggl(
    -\frac{{\cal G}}{{\cal H}}A\D t^2-\frac{2{\cal J}}{{\cal H}}\D t\D r
    +\frac{{\cal F}}{{\cal H}}\frac{\D r^2}{B}
    \notag \\ &\quad 
    +r^2\D\sigma^2
    \biggr),\label{eq:eff_met_1}
\end{align}
with
\begin{align}
    \Omega^2:=\frac{B}{A}\frac{{\cal H}^2}{\sqrt{{\cal FG}+(B/A){\cal J}^2}},
    \label{eq:def:Omegafactor}
\end{align}
and $D_\mu$ is the covariant derivative operator defined
in terms of a connection compatible with $Z_{\mu\nu}$.
Note here that the metric perturbations
have already been expanded in terms of the spherical harmonics
and hence the spherical Laplacian in $Z^{\mu\nu}D_\mu D_\nu$ must be
replaced with its eigenvalue $-\ell(\ell+1)$.
Note also that 
\begin{align}
    \zeta^2(r):={\cal FG}+\frac{B}{A}{\cal J}^2>0
\end{align}
must be imposed in order for the effective metric to be well-defined.
It is easy to see that one has
$Z_{\mu\nu}=\mpl^2 \overline{g}_{\mu\nu}$ in general relativity,
where ${\cal F}={\cal G}={\cal H}=\mpl^2$ and ${\cal J}=0$.
However, $Z_{\mu\nu}$ may not be proportional to $\overline{g}_{\mu\nu}$
in modified gravity. This fact itself has already been known
in the context of the Horndeski
theory~\cite{Minamitsuji:2015nca,Tanahashi:2017kgn,Benkel:2018qmh}.

We introduce a new time coordinate $\tau$
defined by
\begin{align}
    \D \tau = \D t+\frac{{\cal J}}{A{\cal G}}\D r.
\end{align}
Using $\tau$, the effective metric~\eqref{eq:eff_met_1}
can be written in a diagonal form as
\begin{align}
    Z_{\mu\nu}\D x^\mu\D x^\nu &=\Omega^2\left(
    -\frac{{\cal G}}{{\cal H}}A\D \tau^2+\frac{\zeta^2}{{\cal GH}}
    \frac{\D r^2}{B}
%    \notag \\ &\quad 
    +r^2\D\sigma^2  
    \right).
\end{align}

It is sometimes more convenient to work
in the conformally related effective metric $\widetilde Z_{\mu\nu}$
defined as
\begin{align}
    \widetilde Z_{\mu\nu}=\Omega^{-2}Z_{\mu\nu}.
\end{align}
In the tilded frame, Eq.~\eqref{eq:RWZ01} is written as
\begin{align}
    \widetilde Z^{\mu\nu}\widetilde D_\mu\widetilde D_\nu
    \left(\frac{\widetilde\chi}{r}\right) - \left[\frac{V}{{\cal H}}
    +\frac{\widetilde Z^{\mu\nu}\widetilde D_\mu\widetilde D_\nu\Omega}{\Omega}
    \right]
    \frac{\widetilde\chi}{r}=0,
    \label{eq:RWZ02}
\end{align}
where $\widetilde \chi:=\Omega r \chi$
and $\widetilde D_\mu$ is the covariant derivative operator defined
in terms of a connection compatible with $\widetilde Z_{\mu\nu}$.

Defining the generalized tortoise coordinate by
\begin{align}
    \D r_*=\frac{\zeta}{{\cal G}\sqrt{AB}}\D r,
\end{align}
Eq.~\eqref{eq:RWZ02} can further be rewritten in a more familiar form as
\begin{align}
    \left(-\partial_\tau^2+\partial_{r_*}^2-\widetilde V\right)\widetilde\chi = 0,
    \label{eq:RWeq-final}
\end{align}
where
\begin{align}
    \widetilde V&=
    \frac{{\cal G}A}{{\cal H}}\Biggl\{
    \frac{(\ell+2)(\ell-1)}{r^2}
    %\notag \\ &\quad 
    +\Omega r \left[
    \frac{{\cal G}\sqrt{AB}}{\zeta}
    \left(\frac{1}{r\zeta^{1/2}}\right)'
    \right]'\Biggr\}.
    \label{def:effective_potential}
\end{align}
This generalizes the Regge-Wheeler equation known in general relativity~\cite{Regge:1957td}
to higher-order scalar-tensor theories.
In Appendix~\ref{app:source-RW}, we extend
the main result of this section to include
the energy-momentum tensor of matter and derive
the generalized Regge-Wheeler equation with a matter source term

So far we have focused on the modes with $\ell\ge 2$.
The dipole ($\ell=1$) mode must be treated separately,
but here we only comment that the dipole perturbation
corresponds to adding a slow rotation, as has been already discussed in detail in the previous
literature~\cite{Kobayashi:2012kh,Ogawa:2015pea,Takahashi:2016dnv,Takahashi:2019oxz}.

\subsection{Propagation Speed}

In theories described by the action~\eqref{eq:action1},
the propagation speed of gravitational waves differs in general from
the speed of light. In light of the constraint from
GW170817~\cite{TheLIGOScientific:2017qsa,Monitor:2017mdv,GBM:2017lvd},
let us identify the subclass of scalar-tensor theories
that admits a luminal speed of gravitational waves at least at large $r$.
This weak requirement was also employed in Ref.~\cite{Kuntz:2020yow}
(see, however, Refs.~\cite{Babichev:2017lmw,Tattersall:2018map}).

We assume that the background is given by
\begin{align}
    &A=1+{\cal O}(r^{-1}),
    \quad
    B=1+{\cal O}(r^{-1}),
    \notag \\ &
    \psi'=\psi_\infty'+{\cal O}(r^{-1}),
\end{align}
for large $r$, where $\psi_\infty'$ is a constant.
We then find
\begin{align}
    {\cal F}&=2\left[F_2(X_\infty)+\mu^2A_1(X_\infty)\right]+{\cal O}(r^{-1}),
    \label{eq:Flarger}
    \\
    {\cal G}&=2\left[F_2(X_\infty)-(\psi'_\infty)^2A_1(X_\infty)\right] +{\cal O}(r^{-1}),
    \\
    {\cal H}&=2\left[F_2(X_\infty)+2X_\infty A_1(X_\infty)\right]+{\cal O}(r^{-1}),
    \\
    {\cal J}&=-2\mu\psi_\infty'A_1(X_\infty)+{\cal O}(r^{-1}),
    \label{eq:Jlarger}
\end{align}
where $X_\infty:=[\mu^2-(\psi_\infty')^2]/2$.
Thus, if one has
\begin{align}
    A_1(X_\infty)=0,\label{condtion:A1cGW}
\end{align}
Eq.~\eqref{eq:RWeq-final} reduces to
$[-\partial_t^2+\partial_r^2-\ell(\ell+1)/r^2]\widetilde\chi\simeq 0$
for large $r$, rendering luminal propagation of gravitational waves
sufficiently away from a black hole. Note that
$F_{3}$ and $B_I$ appear only in the ${\cal O}(r^{-1})$ terms
in Eqs.~\eqref{eq:Flarger}--\eqref{eq:Jlarger}.

\subsection{Horizons for Photons and Gravitons}

Suppose that $r_h$ is the location of the horizon in the metric $\overline{g}_{\mu\nu}$
and the metric components are expanded as
\begin{align}
    A(r)=\sum_{n=1}\alpha_n\epsilon^n ,\quad 
    B(r)=\sum_{n=1}\beta_n\epsilon^n,\label{eq:expand-A-B}
\end{align}
where $\epsilon:=r/r_h-1>0$.
We assume that $X$ is regular at the horizon, so that
$X$ is of the form
\begin{align}
    X=X_h+\sum_{n=1}X_n\epsilon^n.
\end{align}
Accordingly, one has 
\begin{align}
    \psi'=\frac{\mu}{\sqrt{\alpha_1\beta_1}}\frac{1}{\epsilon}+\sum_{n=0}\gamma_n\epsilon^n.
    \label{eq:expand-dpsi}
\end{align}
Note that $\psi'$ diverges as $r\to r_h$, but this is not problematic.
See the comment below Eq.~\eqref{eq:psi-null}.
Substituting Eqs.~\eqref{eq:expand-A-B}--\eqref{eq:expand-dpsi}
to Eqs.~\eqref{coeff:a1}--\eqref{coeff:a4}, we find, in the vicinity of the horizon,
\begin{align}
    &{\cal F}=-\frac{d_0}{\epsilon}-d_1+{\cal O}(\epsilon),\quad
    {\cal G}=\frac{d_0}{\epsilon}+d_2+{\cal O}(\epsilon),
    \notag \\ &
    {\cal H}=d_3+{\cal O}(\epsilon),\quad
    \sqrt{\frac{B}{A}}{\cal J}=\frac{d_0}{\epsilon}+\frac{d_1+d_2}{2}+{\cal O}(\epsilon),
\end{align}
and hence $\zeta={\rm const}+{\cal O}(\epsilon)$, where
\begin{align}
    d_0&=-\frac{2\mu^2}{\alpha_1}A_1(X_h)+\frac{2\mu}{r_h}\sqrt{\frac{\beta_1}{\alpha_1^3}}
    \bigl[(\alpha_1X_h-2\mu^2)B_2(X_h)
    \notag \\ & \quad 
    -3\mu^2B_3(X_h)+\mu^2X_1B_6(X_h)\bigr],
\end{align}
while the explicit expressions for $d_1$, $d_2$, and $d_3$ are more involved.
Hereafter we will consider the case where $d_0$ is nonvanishing.
Thus, at $r\simeq r_h$,
\begin{align}
    \Omega\simeq {\rm const},
    \quad 
    \widetilde Z_{\tau\tau} \simeq {\rm const},
    \quad 
    \widetilde Z_{rr}\simeq {\rm const},
\end{align}
which shows that
nothing special happens in the effective metric
at the horizon of the metric $\overline{g}_{\mu\nu}$.
In particular, this fact implies that $r=r_h$
is not an appropriate place to impose the inner boundary conditions
when solving the Regge-Wheeler equation~\eqref{eq:RWeq-final}.
Rather, the form of the effective metric implies that
a possible appropriate boundary will be $r=r_g$, where ${\cal G}(r_g)=0$.
To see this more explicitly, let us study some concrete examples.

The first example is given by the special case of
the solution in Sec.~\ref{sec:Background},
with $A_1(X_0)\neq 0$ and $B_1(X_0)=0$.
Essentially the same solution is also studied in Ref.~\cite{Takahashi:2019oxz}.
This does not satisfy Eq.~\eqref{condtion:A1cGW}, but
is a good illustrative example.
We have
\begin{align}
    {\cal G}=2F_2(X_0)\cdot \frac{1-r_g/r}{1-r_h/r},
    \quad 
    {\cal H}=2F_2(X_0)(1+{\cal A}),
\end{align}
where
\begin{align}
    r_g:=(1+{\cal A})r_h,\quad {\cal A}:=\frac{2X_0A_1(X_0)}{F_2(X_0)},
\end{align}
and we assume that $F_2(X_0)>0$ and $1+{\cal A}>0$.
The conformal factor is a nonvanishing constant, $\Omega^2=2F_2(X_0)(1+{\cal A})^{3/2}$,
and the components of the (tilded) effective metric are given by
\begin{align}
    \widetilde Z_{\tau\tau}=-\frac{1-r_g/r}{1+{\cal A}},
    \quad 
    \widetilde Z_{rr}=\frac{1}{1-r_g/r},
\end{align}
which shows that the horizon of the effective metric is at $r=r_g\,(\neq r_h)$.
In this case, the generalized tortoise coordinate is given by
$r_*=(1+{\cal A})^{1/2}[r+r_g\ln(r/r_g-1)]$
and the potential in Eq.~\eqref{eq:RWeq-final} reads
\begin{align}
    \widetilde V=\frac{1-r_g/r}{1+{\cal A}}\left[
    \frac{\ell(\ell+1)}{r^2}-\frac{3r_g}{r^3}
    \right].
\end{align}
Aside from the constant factor of $(1+{\cal A})^{-1}$, this
coincides with the well-known potential in the Regge-Wheeler equation
in general relativity with the horizon at $r=r_g$.

In this example, ${\cal G}$ is singular at $r=r_h$.
One also notices that ${\cal G}<0$ for $r_g<r<r_h$ if ${\cal A}<0$.
However, the effective metric and the potential
do not depend on $r_h$ explicitly and are free from any pathologies.
In particular, the sign of ${\cal G}$ does not directly related to
the stability of the solution. Indeed, it is now clear that
the above solution is stable provided that $F_2(X_0)>0$ and $1+{\cal A}>0$
are satisfied.

The second example is again the special case of the
solution in Sec.~\ref{sec:Background},
but now with $A_1(X_0)=0$ and $B_1(X_0)\neq 0$.
In this case, we have
\begin{align}
    {\cal G}=2F_2(X_0)\cdot\frac{f(r)}{1-r_h/r},\quad {\cal H}=2F_2(X_0),
\end{align}
where
\begin{align}
    f(r)=1- \frac{r_{h}}{r} + {\cal B} \left ( \frac{r_{h}}{r} \right ) ^{5/2},
    \quad 
    {\cal B}:= \frac{81}{2} \frac{\mu^{3}}{r_{h}}\frac{B_{1}(X_{0})}{F_{2}(X_{0})}.
\end{align}
The conformal factor is given by
\begin{align}
    \Omega^2=\frac{2F_2(X_0)}{g^{1/2}(r)},
\end{align}
and the (tilded) effective metric reduces to
\begin{align}
    \widetilde Z_{\tau\tau}=-f(r),\quad \widetilde Z_{rr}=\frac{g(r)}{f(r)},
    \label{eq:eff_met_ex2}
\end{align}
where 
\begin{align}
    %f(r)=&1- \frac{r_{h}}{r} + {\cal B} \left ( \frac{r_{h}}{r} \right ) ^{5/2} \\
    g(r)=&1-{\cal B} \left ( \frac{r_{h}}{r} \right ) ^{3/2}.
\end{align}
We see that the horizon of the effective metric is
at $r=r_g\neq r_h$, where $f(r_g)=0$.

\begin{figure}[tb]
    \begin{center}
            \includegraphics[keepaspectratio=true,height=50mm]{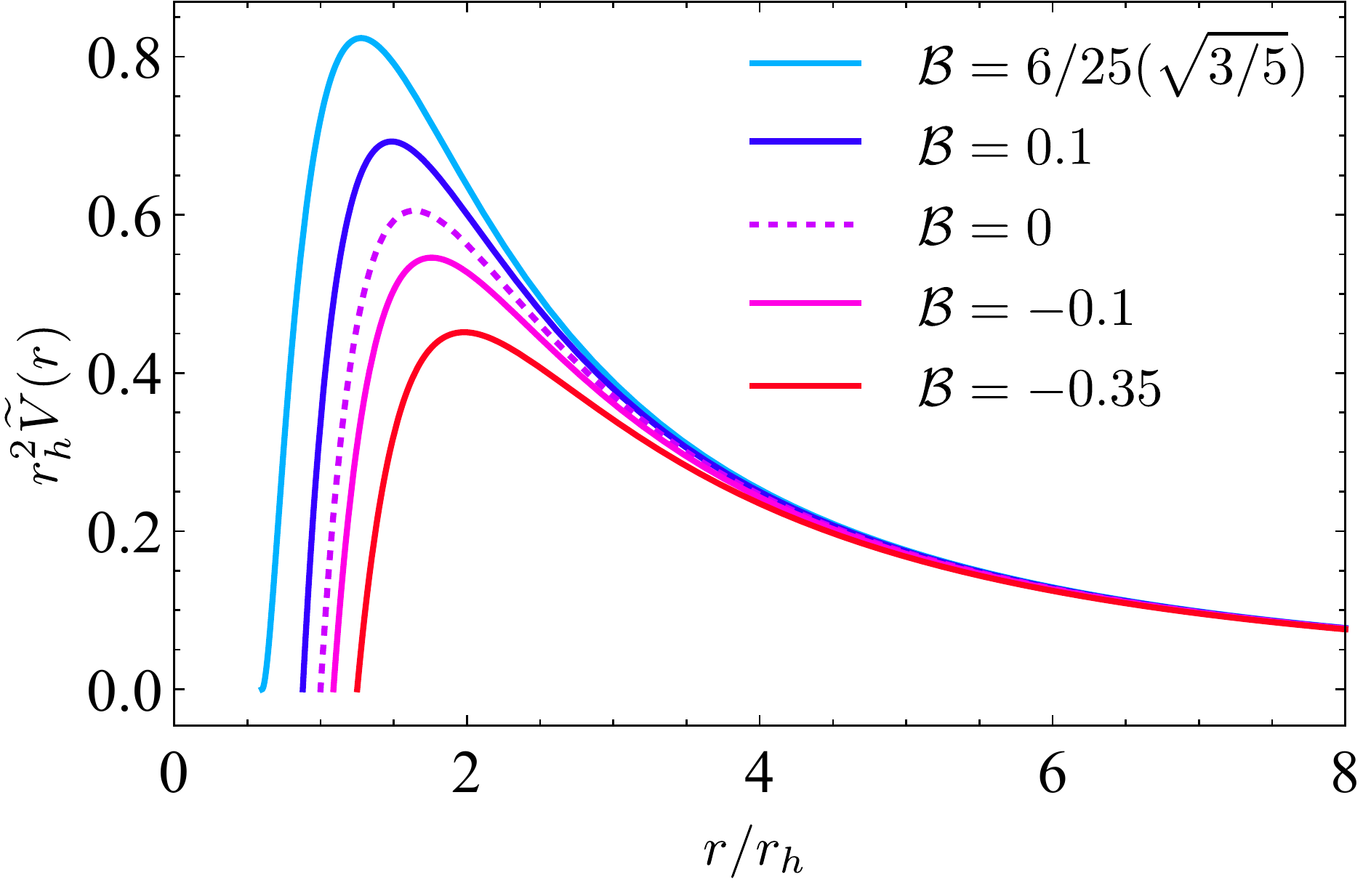}
    \end{center}
      \caption{Potential $\widetilde V$ with $\ell=2$ as a function of $r/r_h$.
  	}
      \label{fig:potential.pdf}
  \end{figure}

Let us investigate the structure of the effective metric~\eqref{eq:eff_met_ex2}
in more detail. For ${\cal B}>6/25(\sqrt{3/5}) \,(\simeq 0.186)$,
$f$ has no zeros, while $g=0$ at $r={\cal B}^{2/3}r_h$. We are not
interested in this case. For $0<{\cal B}\le 6/25(\sqrt{3/5})$,
we have $f=0$ at $r=r_g<r_h$. In this case,
$g$ remains positive outside the horizon of the
effective metric, but $g=0$ occurs at $r={\cal B}^{2/3}r_h<r_g$.
Finally, for ${\cal B}<0$, we have
$f=0$ at $r=r_g>r_h$ and $g$ is always positive for $r>0$.
Therefore, in the latter two cases the solution has an outer horizon
of the effective metric at $r=r_g$. It is straightforward to write
the potential $\widetilde V$, but the expression is messy.
The shape of the potential is shown
for different values of ${\cal B}$ in Fig.~\ref{fig:potential.pdf}.
One can check that $r_*\to-\infty$ as $r\to r_g$.

%-----------------------------------------------%
%-----------------------------------------------%

\section{Quasi-Normal Modes}

In this section, we compute the QNMs of the
second example of the previous section.
Quasi-normal modes in the Horndeski theory
have been studied in the case of the
Schwarzschild background with a constant scalar field~\cite{Tattersall:2018nve}
and a nearly Schwarzschild background with
a nearly constant scalar field~\cite{Tattersall:2019nmh}.

\begin{figure}[tb]
    \begin{center}
            \includegraphics[keepaspectratio=true,height=52mm]{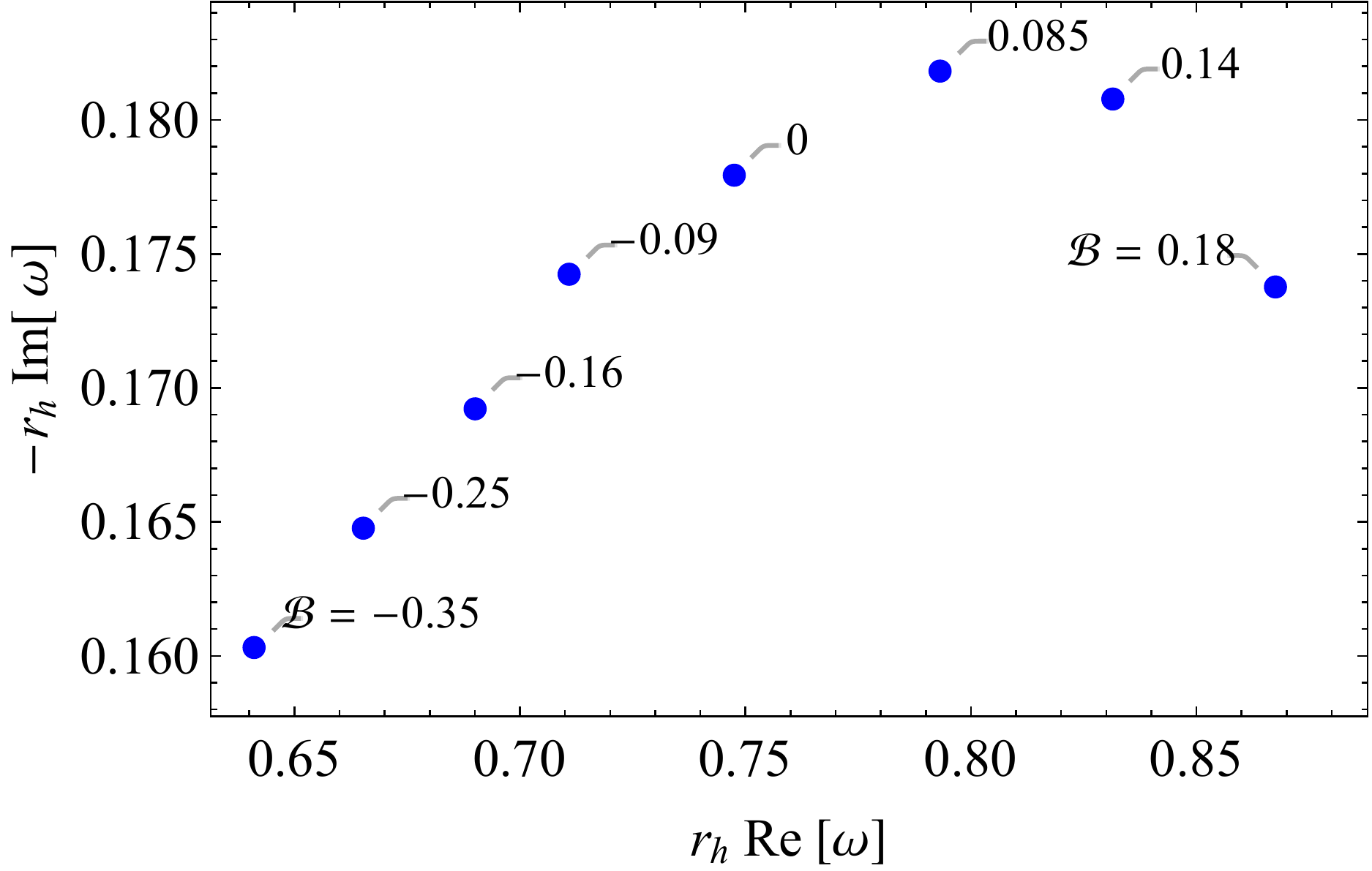}
    \end{center}
      \caption{Lowest overtone quasi-normal frequencies for
      $\ell=2$ and some representative values of ${\cal B}$.
  	}
      \label{fig:QNM.pdf}
  \end{figure}

We assume the time dependence of the master variable
as $\widetilde\chi=Q(r)e^{-i\omega\tau}$ and solve
\begin{align}
    \frac{\D^2Q}{\D r_*^2}+\left[\omega^2-\widetilde V(r)\right]Q=0,
\end{align}
where $\widetilde V$ is the potential obtained from the second example
of the background solutions in the previous section,
which is characterized by the dimensionless
parameter ${\cal B}$ (Fig.~\ref{fig:potential.pdf}).
The boundary conditions for $Q$ are given by
\begin{align}
    Q\propto  
    \begin{cases}
    e^{+i\omega r_*} \quad & r_*\to\infty\quad (r\to\infty) \\
    e^{-i\omega r_*} \quad & r_*\to-\infty \quad (r\to r_g)
  \end{cases}.
\end{align}
Note again that the inner boundary is located at $r=r_g$ rather than at $r=r_h$.
In order to obtain the QNMs, we employ direct numerical integration.\footnote{Taking
${\cal B}$ as a small expansion parameter, one may write the potential
as $\widetilde V=V_{\rm GR}+\delta V$, where $V_{\rm GR}$ is the
Regge-Wheeler potential in general relativity and $\delta V$ is a small correction.
In the present case, $\delta V$ contains fractional powers of $r$, which
hinders us from using the convenient formalism of Ref.~\cite{Cardoso:2019mqo}.}
The lowest overtone quasi-normal frequencies for $\ell=2$
are given in Fig.~\ref{fig:QNM.pdf},
showing how the frequencies depend on
the modified gravity parameter ${\cal B}$.

%-----------------------------------------------%
%-----------------------------------------------%

\section{Conclusions}

In this paper, we have studied odd parity perturbations of
black holes with linearly time-dependent scalar hair in shift-symmetirc
scalar-tensor theories.
Due to the time dependence of the scalar field background,
the effective-field-theory (EFT)
approach~\cite{Kase:2014baa,Franciolini:2018uyq,Kuntz:2020yow}
can not be applied straightforwardly to the present case.
Therefore, we have started from a general covariant action
that is most similar to the action of cubic degenerate higher-order
scalar-tensor theories~\cite{BenAchour:2016fzp}
and derived the general quadratic action for odd parity perturbations
without imposing the degeneracy conditions.
The degeneracy conditions are not essential for
retaining the healthy odd parity perturbations that are not
mixed with the perturbation of the scalar field.
We have thus derived a second-order equation for a single master variable
as a generalization of the Regge-Wheeler equation in general relativity.
Starting from the more general action, we have arrived at
qualitatively the same results as the previous
ones~\cite{Ogawa:2015pea,Takahashi:2016dnv,Takahashi:2019oxz},
showing that no new terms appear in the quadratic action for
odd parity perturbations. Our generalized Regge-Wheeler equation
can be used in a wide class of scalar-tensor theories
such as U-degenerate theories~\cite{DeFelice:2018ewo} and
scordatura theories~\cite{Motohashi:2019ymr}.

We have also refined the stability conditions explored in the
previous literature~\cite{Takahashi:2019oxz}. The previous conditions were actually
sufficient conditions, and we have argued that one of the conditions
is not directly related to the stability.

As another application of our results, we have computed the
quasi-normal modes of a certain nontrivial black hole solution.
In doing so, we have demonstrated that
it is important to identify the correct location of
the inner boundary by inspecting the effective metric for gravitons.

It would be interesting to extend the present analysis to
the even parity sector, which would be much more involved due to
its higher derivative nature.
It would also be interesting to perform a complementary analysis
based on the EFT approach along the line of~\cite{Franciolini:2018uyq,Kuntz:2020yow}.

%--- Acknowledgments ---%--- Acknowledgments ---%--- Acknowledgments ---%
\acknowledgments
We are grateful to Masashi Kimura for sharing his Mathematica code
for calculating quasi-normal modes.
We also thank Hayato Motohashi and Kazufumi Takahashi
for their comments on the manuscript.
The work of KT was supported by
the Rikkyo University Special Fund for Research (SFR).
The work of TK was supported by
JSPS KAKENHI Grant Nos.~JP20H04745 and~JP20K03936.
%--- Acknowledgments ---%--- Acknowledgments ---%--- Acknowledgments ---%

%-----------------------------------------------------------------%

\appendix

\section{Generality of the Quadratic Lagrangian}

Starting from the action~\eqref{eq:action1},
we have shown in the main text that the quadratic Lagrangian for
the odd parity modes is given by Eq.~\eqref{eq:quadratic_Lagrangian}.
Actually, one can show that more general scalar-tensor theories
lead to the quadratic Lagrangian for the odd parity modes
having the same structure as Eq.~\eqref{eq:quadratic_Lagrangian}
as long as the equation of motion for gravitational-wave
degrees of freedom remains of second order.

For example, one may add to the action~\eqref{eq:action1}
\begin{align}
    \widetilde F_3(X)R\Box\phi,
\end{align}
to consider a general derivative coupling of the form
$F_3G_{\mu\nu}\phi^{\mu\nu}+\widetilde F_3R\Box\phi%
=F_3R_{\mu\nu}\phi^{\mu\nu}+(\widetilde F_3-F_3/2)R\Box\phi$.
This only shifts the coefficients as
\begin{align}
    {\cal F,G,H}&\to {\cal F,G,H}+\left[
    \frac{B\psi'}{r}-\frac{(AX)'}{A\psi'}
    \right]\widetilde F_3,
    \\
    {\cal J}& \to {\cal J} ,
\end{align}
and does not give rise to any new terms in Eq.~\eqref{eq:quadratic_Lagrangian}.

Similarly, one may also add terms quartic in second derivatives of $\phi$
such as
\begin{align}
    C_1(X)\phi_{\mu\nu}\phi^{\nu\rho}\phi_{\rho\lambda}\phi^{\lambda\mu} ,
    \quad 
    C_2(X)(\Box\phi)^4,\quad \cdots.
\end{align}
One can verify by direct computation that
such quartic terms merely shift the coefficients without altering the
structure of the Lagrangian~\eqref{eq:quadratic_Lagrangian}
or have no contribution to the odd parity sector.

We thus conclude that the form of the Lagrangian~\eqref{eq:quadratic_Lagrangian}
is generic to scalar-tensor theories in which gravitational-wave
degrees of freedom obey a second-order equation of motion.

\section{Sourced Regge-Wheeler Equation}\label{app:source-RW}

In this appendix,
we generalize our main result to include the source term,
which has not been considered in the previous similar
studies~\cite{Kobayashi:2012kh,Ogawa:2015pea,Takahashi:2016dnv,Takahashi:2019oxz}.
Assuming that matter is minimally coupled to gravity,
the source term can be obtained from
\begin{align}
    S_{\rm source}=\frac{1}{2}\int \D^4x\sqrt{-\overline{g}}h^{\mu\nu}T_{\mu\nu},
    \label{intaction01}
\end{align}
where $T_{\mu\nu}$ is the matter energy-momentum tensor.
Similarly to the metric perturbations, the odd parity part of
the energy momentum tensor can also be expanded as
\begin{align}
    T_{t\theta}&=-\frac{1}{\sin\theta}\partial_\varphi
    \sum_{\ell=2}^\infty\sum_{m=-\ell}^{\ell} S_0^{(\ell m)}
    (t,r)Y_{\ell m},\label{eq:Tmn_def_1}
    \\
    T_{t\varphi}&=\sin\theta\partial_\theta
    \sum_{\ell=2}^\infty\sum_{m=-\ell}^{\ell} S_0^{(\ell m)}
    (t,r)Y_{\ell m},
    \\
    T_{r\theta}&=-\frac{1}{\sin\theta}\partial_\varphi
    \sum_{\ell=2}^\infty\sum_{m=-\ell}^{\ell} S_1^{(\ell m)}
    (t,r)Y_{\ell m},
    \\
    T_{r\varphi}&=\sin\theta\partial_\theta
    \sum_{\ell=2}^\infty\sum_{m=-\ell}^{\ell} S_1^{(\ell m)}
    (t,r)Y_{\ell m},
    \\
    T_{\theta\theta}&=\frac{2}{\sin\theta}\left(
    \partial_\theta\partial_\varphi -\cot\theta \partial_\varphi\right)
    \sum_{\ell=2}^\infty\sum_{m=-\ell}^{\ell} S_2^{(\ell m)}
    (t,r)Y_{\ell m},
    \\
    T_{\theta\varphi}&=\left(\frac{1}{\sin\theta}\partial_\varphi^2
    +\cos\theta\partial_\theta-\sin\theta\partial_\theta^2\right)
    \notag \\ 
    &\quad \times
    \sum_{\ell=2}^\infty\sum_{m=-\ell}^{\ell} S_2^{(\ell m)}
    (t,r)Y_{\ell m},
    \\
    T_{\varphi\varphi}&=-2\sin\theta\left(
    \partial_\theta\partial_\varphi -\cot\theta \partial_\varphi\right)
    \sum_{\ell=2}^\infty\sum_{m=-\ell}^{\ell} S_2^{(\ell m)}
    (t,r)Y_{\ell m}.
    \label{eq:Tmn_def_4}
\end{align} % minus of Maggiore's definition
The conservation of the matter energy-momentum tensor,
$\nabla_\nu T^{\mu\nu}=0$, yields
\begin{align}
    &-\frac{\dot S_0^{(\ell m)}}{A} + 
    \frac{\sqrt{B/A}}{r^2}\left(r^2\sqrt{AB}S_1^{(\ell m)}\right)'
    \notag \\ &
    +\frac{(\ell-1)(\ell+2)}{r^2}S_2^{(\ell m)}=0.
\end{align}

It is straightforward to perform the angular integrations
in Eq.~\eqref{intaction01} to obtain
\begin{align}
S_{\rm source}&=-\sum_{\ell=2}^\infty\sum_{m=-\ell}^\ell
\frac{\ell(\ell+1)}{2}
\notag \\ &\quad \times
\int\D t \D r\left(
\frac{h_0^*S_0}{\sqrt{AB}}-\sqrt{AB}h_1^*S_1+{\rm c.c.}
\right),
\end{align}
where we omitted the labels $(\ell m)$ from $S_0$ and $S_1$.
This is the source action for the odd mode perturbations
(see also Ref.~\cite{Kuntz:2020yow}).
We add the above source action to the gravitational part of the
action~\eqref{gravaction2}. Then, Eqs.~\eqref{eq:h0eq} and~\eqref{eq:h1eq}
are generalized to
\begin{align}
    &a_1h_0+(a_3\chi)'+\frac{2a_3}{r}\chi +\frac{1}{2}a_4h_1=\frac{\ell(\ell+1)}{2\sqrt{AB}}S_0,
    \\
    &a_2h_1-a_3\dot \chi +\frac{1}{2}a_4h_0=-
    \frac{\ell(\ell+1)}{2}\sqrt{AB}S_1,
\end{align}
Solving these equations for $h_0$ and $h_1$
and removing $h_0$ and $h_1$ from the quadratic Lagrangian,
we see that the Lagrangian~\eqref{eq:Lagrangian-chi2} is
generalized to include the source as 
\begin{align}
    {\cal L}_{\ell m,{\rm total}}^{(2)}= {\cal L}_{\ell m}^{(2)}-
    \frac{\ell(\ell+1)}{4(\ell-1)(\ell+2)}\sqrt{\frac{B}{A}}
    \left(\chi^*S_{\rm odd}+{\rm c.c.}\right),
\end{align}
where ${\cal L}_{\ell m}^{(2)}$ in the right-had side
is the same Lagrangian as the one defined as Eq.~\eqref{eq:Lagrangian-chi2}
and
\begin{align}
    S^{(\ell m)}_{\rm odd}(t,r)&:=
    2{\cal H}\left(\frac{{\cal G}}{\zeta^2}S_0^{(\ell m)}\right)'
    -\frac{2{\cal FH}}{\zeta^2}\dot S_1^{(\ell m)}
    \notag \\ &\quad 
    -\frac{2{\cal HJ}}{A\zeta^2}\dot S_0^{(\ell m)}
    -2{\cal H}\left(\frac{B{\cal J}}{\zeta^2}S_1^{(\ell m)}\right)'
    .
\end{align}
Now Eq.~\eqref{eq:RWZ01} with the source term reads
\begin{align}
    {\cal H}\Omega^2 Z^{\mu\nu}D_\mu D_\nu\chi -V\chi =S_{\rm odd},
\end{align}
and, accordingly, Eq.~\eqref{eq:RWeq-final} with the source term is given by
\begin{align}
    \left(-\partial_\tau^2+\partial_{r_*}^2-\widetilde V\right)\widetilde\chi = 
    \frac{{\cal G}r\sqrt{AB}}{{\cal H}\zeta^{1/2}}S_{\rm odd}.
\end{align}
This is the generalization of the sourced Regge-Wheeler equation.

%-----------------------------------------------------------------%
\bibliography{refs}
\bibliographystyle{JHEP}
%-----------------------------------------------------------------%
\end{document}